\documentclass[12pt, a4paper]{article}
\usepackage{amsthm, amsmath, amsfonts, emlines, epsf}

\def\ds{\displaystyle}
\def\bea{\begin{eqnarray}}
\def\eea{\end{eqnarray}}
\def\be{\bea\begin{array}{c}\ds}
\def\ee{\end{array}\eea}

\def\sm{\sigma^\mu}
\def\sbm{\bar{\sigma}^\mu}
\def\gm{\gamma^\mu}
\def\pq{\psi_q}
\def\pt{\psi_{\tilde{q}}}

\def\pbq{\bar{\psi}_{\tilde{q}}}

\begin{document}
\vspace*{2cm}
\setcounter{footnote}{0}
\def\thefootnote{\fnsymbol{footnote}}
\begin{center}
\hfill ITEP-TH-88/02\\
\vspace{0.4in}
{\LARGE\bfseries Comments on BPS Bound State ``Decay"}
\end{center}
\vspace{0.2in}
\centerline{{\large Anatoly Dymarsky\footnote{
dymarsky@gate.itep.ru
}
and Dmitry Melnikov\footnote{
melnikov@gate.itep.ru}}
}
\vspace{0.3in}
\begin{center}
\large
\it Department of Physics at Moscow State University
\\ Vorobjevy Gory, 119899 Moscow, Russia
\vspace{0.3in}
\\ Institute for Theoretical and Experimental Physics
\\ B.Cheremushkinskaya 25, 117259 Moscow, Russia\\
\vspace{0.3in}
\rm March 2003
\end{center}

\bigskip

\begin{abstract}
{\footnotesize In ${\cal{N}}=2$ SYM  the charges of a
BPS state shift under the transition to a dual
description of the theory. In particular, in the theory with matter duality transformations
may convert a bound state to unbound one as was predicted by
Seiberg and Witten from considerations of the monodromies around
the moduli space singularities. A physical mechanism of such a
behavior on the semiclassical level could be established
explicitly through consideration of soliton-fermion classical
field configurations. The problem reduces to the investigation of fermion spectrum in a slowly varying background monopole field. Behavior of the solutions to classical
equations allows one to observe the BPS bound state ``decay" \it in
vivo.}
\end{abstract}


\bigskip

\section{Introduction}
One of the greatest recent successes of Supersymmetry applications was the
evaluation of low energy effective action for ${\cal{N}}=2$
supersymmetric Yang-Mills theories. This became possible, since Supersymmetry imposed strong restrictions on the theory, and in particular, on the form of the action. It turns out, that the only thing one is left to do then, instead of computing complicated path integrals, is to develop an intuition of working with the complex space describing physical vacua of the theory. This was done by Seiberg and Witten in \cite{SW1},\cite{SW2}.

This solution of the supersymmetric Yang-Mills theories gave a lot of material for the future investigations.
The main purpose of this work is to study the behavior of a certain class of supersymmetric states, namely BPS states, under the adiabatic variation of low energy
 moduli of the theory. BPS states are of crucial importance in SUSY theories, since their spectrum can be easily found. We shall focus on the ${\cal{N}}=2$ $SU(2)$ SYM theory with matter and mainly consider the BPS
  states in the sector with nontrivial magnetic charge. As we are going to demonstrate, under certain
variation of the moduli, localized BPS state become delocalized. Predictions of such a behavior could be
easily obtained from the exact solution of the theory. They are actually presented in the original paper \cite{SW2}. However,
 the explanation given there uses only the general arguments (which we briefly review in the next section)
 and does not provide the explicit mechanism of such unusual behavior. It was also noticed there that treating coupling
 constant small, one could examine this phenomena in details through considering classical solutions to equations of
 motion. Our goal here is to fulfill this program and to demonstrate how this mechanism  explicitly  works.

According to Seiberg and Witten, to obtain the dual description of
the theory, one should consider the variation of moduli, which
corresponds to a closed loop on the moduli space around singular
point, where the low-energy description fails. If this loop is
topologically  trivial, the corresponding transformation is
trivial and we obtain the same theory description. But if we
encircle singular point, we come to another description of the
same theory. As we know from the famous example of
electric-magnetic duality due to Montonen and Olive \cite{MO},
duality transformation may interchange the roles of electric and
magnetic charges. Something similar happens in the above case.
Seiberg and Witten proposed that the Montonen-Olive duality is
only a special case of more general group of duality
transformations.

In the present work we shall consider the theory only in the
vicinity of one type of singularities. These are the points, where
the  quark  become massless. Let us focus on the sector with unit
magnetic charge. The topic of our main interest is the combined
state, consisted of monopole and fermion. Under certain conditions
these two objects may coexist as a bound state, i.e. the
configuration with localized energy. The set of points on the
moduli space, where delocalization necessarily occurs, is called
the Curve of Marginal Stability (CMS). We shall refer to these
delocalization as to a ``decay"\footnote{We use a terminology of \cite{Bilal-Ferrari}, where similar physics was studied.} of a bound state.

It happens that the massless quark singularity belongs to CMS .
Thus, one will intersect the CMS twice moving along the small
closed loop around it. At some part of the loop one will have the
monopole-fermion bound state, at the other, this state will be
unbounded. On the other hand since closing the loop, the theory
description may switch to the dual one, one might expect that the
 charges will transform in a nontrivial way, such that unbounded state becomes again bounded. This indeed takes place in the case we are going to consider.

As was mentioned above, we are going to investigate the process of bound state ``decay" or delocalization by solving the classical equations of motion near the singular point, treating this region of the moduli space  semiclassically. We analyze the solutions to the equations of motion and
determine the energy spectrum of the fermion states in the field of
monopole. We also derive the restrictions on existence of the bound state.
Collecting all pieces together, the shift of charges due to nontrivial
duality transformation, predicted by Seiberg-Witten simple arguments, could be explicitly recovered.

The paper is organized as follows: we give more explicit grounds
to the speculations above and present a brief introduction to the
Seiberg-Witten method in the Section 2. The reader familiar with
the topic may skip this introduction. In the Section 3 we solve
the classical equations for fermion-monopole system and analyze
the solution, thus, describing the bound state decay
semiclassically. Considering the transformation of the fermion
spectrum during encircling singularity we reproduce the charge
shift predicted by Seiberg and Witten. Section 4 is concluding. We
also give there several brief comments on the physics underlying
the phenomenon.

This work has several  common points with previous results and we
refer the interested reader for the following papers.  The
conditions for existing of the bound state were first derived in
\cite{HEN}. The notion of BPS state decay was originally
introduced in the works by Bilal and Ferrari \cite{Bilal-Ferrari},
where this phenomenon was first derived. See \cite{Semenoff}
and \cite{Ferrari}, and references therein. Finally, the appropriate CMS was studied in \cite{VainsteinR}.


\section{Prehistoric}

\subsection{Discussion of Seiberg-Witten Theory}
In the two works of 1994 Seiberg and Witten managed to evaluate
the exact low energy effective action for both pure ${\cal{N}}=2$
Super Yang-Mills theory \cite{SW1} and ${\cal{N}}=2$ Super
Yang-Mills theory with matter \cite{SW2}. The method they applied was based on investigation of symmetries of the
action rather than on evaluating functional integrals. It was well
known that supersymmetry imposes severe restrictions on the form
of the action. Thus, for ${\cal{N}}=2$ SYM theory the action is
fixed up to some locally holomorphic function\footnote{For more
detailed reviews of ${\cal{N}}=2$ and ${\cal{N}}=1$ actions see
for example \cite{reviews}.}:
\be
\label{N=2SYM}
S_{N=2}=\frac{1}{16\pi}\Im {\rm m}\left[ {\rm Tr}\int d^2\theta d^2
\tilde{\theta}~{\mathcal{F}} (\Psi)\right],
\ee
where ${\cal{F}}$ is the mentioned locally holomorphic function called prepotential
and $\Psi$ denotes the ${\cal{N}}=2$ superfield. We stress that only due to supersymmetry
the action could have such a simple form.

Classical form of $\cal{F}$ is fixed by renormalizability to be ${\cal{F}}= 1/2\tau\Psi^2$,
with $\tau$ the complex bare coupling constant\footnote{Note that $\tau={\cal{F}}^{\prime\prime}$.
This relation holds in the quantum case as well.}
$$ \tau= {\theta \over 2\pi} + {4\pi i\over g^2},$$
where $\theta$ is the QCD $\theta$-angle. Rather simple expression
(\ref{N=2SYM}) actually encodes the long action containing vector gauge
field $A_\mu$, two Weyl fermion fields $\psi$ and $\lambda$, and one complex
scalar field $A$ in the adjoint representation of some gauge group $G$:
\be
\label{N=2fields}
S_{YM}= \int d^4x~\frac{1}{g^2}\biggl(-\frac{1}{4}F_{\mu\nu}F^{\mu\nu} +
\frac{g^2\theta}{32\pi^2} F_{\mu\nu} \tilde{F}^{\mu\nu}-i\lambda\sm D_\mu
\bar{\lambda}+\frac{1}{2}D^2 + (D_\mu A)^\dagger D_\mu A -\\
i\bar{\psi}\sbm D_\mu\psi - D[A^\dagger,A]-i\sqrt{2}[\lambda,\psi]A^\dagger -
i\sqrt{2}[\bar{\lambda},\bar{\psi}]A+F^\dagger F\biggr).
\ee
As could be seen from (\ref{N=2fields}) the action possesses a certain
potential $V$ arising after integrating out auxiliary fields. In
supersymmetric theories the vacua of the theory are simply the zeroes of
potential:
$$
V=\frac{1}{2}{\rm tr}[A^\dagger,A]^2=0.
$$
Since $A$ belongs to adjoint representation of a gauge group, the
vacua of ${\cal{N}}=2$ theory are described by Cartan elements of
Lie algebra of the group $G$. Further we shall concentrate
exceptionally on the case of SU(2) gauge group. For this case
the vacua corresponds, after appropriate gauge transformation, to
values of $A$ along the $\sigma^3$ direction:
$$
\langle A\rangle = {1\over 2}a\sigma^3.
$$
The complex quantity $a$, parameterizes the space of classical
vacua, called the moduli space of the theory.

Seiberg and Witten were searching for the low energy effective action. This action should contain only massless particles.
 Due to Higgs effect most of the fields acquire masses. However the components of non-abelian fields
 along the $\sigma^3$ direction are still massless and therefore the
effective description should be a functional of these fields. Evaluation of the effective action is generally
 highly  complicated task. However in this case supersymmetry simplify the problem essentially,
 imposing restrictions on the quantum action as well. The massless part of the action (\ref{N=2SYM}) fixed by
  supersymmetry in more familiar ${\cal{N}}=1$ notations has the form
\be
\label{N=1 low energy}
S=\frac{1}{16\pi}\Im {\rm m} \left[ \int d^2\theta~ {\mathcal{F}}^{\prime
\prime} (\Phi) W^\alpha W_\alpha + \int d^4
\theta~\Phi^+{\mathcal{F}}^{\prime} (\Phi)\right],
\ee
where $W_\alpha$ and $\Phi$ denote the ${\cal{N}}=1$ vector and chiral
multiplets respectively. Now to find the effective action we must determine the quantum form of the prepotential $\cal{F}$. This could be done by investigating the moduli space of the theory.

Classically moduli space is the complex plane with one fixed point
$a=0$, where  the broken symmetry restores. We shall refer to this point as to the singularity, since there the effective low energy description (\ref{N=1 low energy}) breaks down. In the quantum case there are several additional singularities on the moduli space, where some massive modes become massless.

If we wrote the expression (\ref{N=1 low energy}) in component fields, we
should notice that the coefficient of the kinetic term for scalar field $A$ would be proportional to imaginary part of second derivative of $\cal{F}$. In other words, the metric on the moduli space would be
$$
ds^2=\Im{\rm m}({\mathcal{F}}^{\prime\prime}(a))dad\bar{a}.
$$
Being a harmonic function this metric can be positively definite only in the case of meromorphic function $\cal{F}$, and the singularities of $\cal{F}$ exactly coincide with the mentioned singularities of moduli space.

Now let us introduce new variable $a_D$ which is the Legendre conjugate of $a$,
\be
\label{a_D}
a_D=\frac{\partial {\mathcal{F}}(a)}{\partial a}.
\ee
This variable will naturally arise in the quantum generalization of the central charge formula.
 The central charge appears as the nontrivial commutator of SUSY generators, being a central extension of the algebra:
$$
\left\{Q^I_\alpha,Q^J_\beta\right\}=\sqrt{2}\epsilon^{IJ}\epsilon_{\alpha\beta}Z.
$$
For a given physical state, central charge is connected with the mass of the
state via the so-called Bogomolny inequality \cite{WO}:
$$
M^2\geq 2\left|Z\right|^2. $$ The states that saturate this
inequality are called BPS states \cite{BPS}. The latter protect
one half of supersymmetry and their mass is equal to the central
charge. The spectrum of these states could be always exactly
determined since the supersymmetry preserve the Bogomolny
inequality in the quantum case as well. One needs only to
determine the central charge for corresponding state. The
classical central charge for ${\cal{N}}=2$ SYM was calculated by
Witten and Olive in \cite{WO}. It is generated by vacuum
expectation value of the scalar field and is naturally expressed
in terms of electric and magnetic charges of the physical state.
The quantum formula for the central charge, introduced in
\cite{SW1}, is supposed to be
\be
\label{SYM CC}
Z=n_ea+n_ma_D.
\ee
This formula reveals the reciprocal  character of $a$ and $a_D$. In fact it
could be analyzed from the action (\ref{N=1 low energy}) that Legendre
transformation (\ref{a_D}) generated by ${\cal{F}}$ leads to the same action in terms of
 dual fields ($\Phi_D$, $W^\alpha_D$), with the new coupling $\tau_D=1/\tau$. Obviously the electric and magnetic charges interchanges under this transformation, but the central charge (\ref{SYM CC}), holds invariant, since it is connected to the observable quantity,the mass of the state. This is the famous Montonen-Olive electric-magnetic duality. Seiberg and Witten found that the full duality group is larger, namely all $SL(2,\mathbb{Z})$ transformations leave the theory invariant.

As soon as every point of the moduli space is characterized by a
pair $(a,a_D)$, there is a spectrum of the BPS states attached to
each point. These states are characterized by definite electric and
magnetic charges. Seiberg and Witten proposed to consider the
central charge formula (\ref{SYM CC}) as the product of vector
consisted of $a_D$ and $a$, and a row of the charges $(n_m,n_e)$.
This form reflects the idea that $a_D$ and $a$ belongs to representation of duality group
$SL(2,\mathbb{Z})$ and helps to understand how the duality or monodromies rotate the charges of the physical state.

To determine the monodromies one could consider the low-energy
theory in the vicinity of singularities and calculate the
perturbative one-loop $\beta$-function, with $a$ playing a role of
scale:
\be
\label{bet}
\beta(g) \equiv a \frac{dg^2}{da}
\ee
 The latter could be established from consideration of the spectrum of
the theory at the  singular point. Indeed the prepotential defines
the coupling constant $\tau={\cal{F}}^{\prime\prime}$, that is
${\tau=\partial a_D / \partial a}$. Derived from (\ref{bet}), behavior of $a_D$ as the function of $a$, while encircling the singularity, fixes the monodromy. This will be done in details in the next subsection. Possessing the information about the number of singularities and their monodromies, it is possible to completely restore the $SL(2,{\mathbb{Z}})$ fiber bundle $(a_D,a)$ over the moduli space, and thus, functions
${\cal{F}}$ and $\tau$. This is enough for recovering the explicit
form of the effective action, however, we do not go in this any
further here.

\subsection{Theory with Matter}
Consideration of the theory with matter implies several slight
modifications. On the level of the Lagrangian one adds a term which
corresponds to $N_f$ ${\cal{N}}=2$ hypermultiplets or $N_f$ pairs of
${\cal{N}}=1$ chiral and antichiral multiplets, either in
adjoint or fundamental representation. We shall be interested in
the choice of fundamental matter and the number of flavors being
one. From the ${\cal{N}}=1$ point of view we add one chiral and
one anti-chiral multiplets, each consisted of complex scalar field
and Weyl fermion. The full Lagrangian in ${\cal{N}}=1$ notations has the
form
\be
\label{N=1 matter}
S = \frac{1}{8\pi}{\rm \Im m~Tr}\biggl[\tau\left(\int d^4x~d^2\theta~W^\alpha W_\alpha+2\int
d^4x~d^2\theta d^2\bar{\theta}~\Phi^\dagger e^{-2V}\Phi\right)\biggr] +\\ \\ \ds + \int
d^4x~d^2\theta d^2\bar{\theta}~\left( Q^\dagger_i e^{-2V}Q_i + \tilde{Q}_i e^{2V}
Q^\dagger_i \right) +\\ \\ \ds + \left( \int d^4x~d^2\theta~\sqrt{2}\tilde{Q}_i\Phi Q_i
+ m_i \tilde{Q}_i Q_i + h.c. \right).
\ee
and its matter part in components looks like
\begin{equation*}
\begin{array}{c}
\label{matter components}
S_{\rm matter}=\int d^4x~(D_\mu q)^\dagger D_\mu q + D_\mu \tilde{q}(
D_\mu\tilde{q})^\dagger - i\bar{\pq}\sbm D_\mu \pq + i\pt \sm D_\mu \pbq -
\\ \\ \ds - q^\dagger D q + \tilde{q} D \tilde{q}^\dagger - i\sqrt{2}q^\dagger
\lambda\pq + i \sqrt{2}\tilde{q}\bar{\lambda}\pbq +
i\sqrt{2}\bar{\pq}\bar{\lambda}q - i \sqrt{2}\pt\lambda\tilde{q}^\dagger +
F_{q}^\dagger F_q + F_{\tilde{q}}F^\dagger_{\tilde{q}} + \\ \\ \ds + \bigr[
\sqrt{2}\tilde{q}A F_q + \sqrt{2}F_{\tilde{q}}A q - \sqrt{2}\pt A \pq -
\sqrt{2}\tilde{q}\psi\pq - \sqrt{2}\pt\psi q + \\ \\ \ds +\tilde{q}Fq +
m\tilde{q}F_q + mF_{\tilde{q}}q - m \pt\pq + h.c. \bigl].
\end{array}
\end{equation*}

The matter part of the action possesses $U(N_f)$ global symmetry.
${\cal{N}}=1$ mass term generally breaks this symmetry down to product
of $N_f$ copies of $U(1)$. Therefore the matter fields care
additional global charges with respect to this $U(1)$'s. We might
assume that these charges also contribute to the central extension
of the SUSY algebra, and it is indeed the case: the central charge
formula (\ref{SYM CC}) modifies in the following way:
\be
\label{Matter CC}
Z=n_ea+n_ma_D+\sum\limits_{N_f}\frac{m_i}{\sqrt{2}}S_i,
\ee
where $S_i$'s are the corresponding $U(1)$ global charges.

Below we are going to concentrate on the particular type of
singularities of the moduli space, namely the ones, corresponding to
the massless quarks. If we look at the central charge formula, we
shall see that the BPS states with the charges $(0, -S, S)$
become massless at the point of the moduli space $a_0=m/\sqrt{2}$. In this point the matter fields (quarks and scalars) become massless. This fact could also be easily derived from the classical Lagrangian (\ref{matter components}). Since the spectrum of the low energy effective theory is known, the one-loop $\beta$-function could be calculated in the vicinity of this point:
$$
\beta(g)=\frac{g^3}{8\pi^2},
$$
This $\beta$-function is exact perturbatively since the higher
loop corrections are forbidden by supersymmetry. Furthermore,
since the non-perturbative part is regular  in the vicinity of
singularity we could simply drop it. From the $\beta$-function we
restore the leading logarithmic part of coupling $\tau$ as
the function of the scale $a$ and next, the leading order of the function $a_D$, which has the following form near the point $a_0$:
\be
\label{a_D singular}
a_D\simeq c-\frac{i}{2\pi}(a-a_0)\ln(a-a_0).
\ee

For the theory with matter, the central charge  (\ref{Matter
CC}) is represented respectively as the product of the row of
charges $(n_m,n_e,S_f)$ and a column, consisted of $a_D$, $a$, and
$m$. The expansion (\ref{a_D singular}) allows one to investigate
the monodromies of $a_D$ around the singularity. Performing a
closed contour around $a_0$ in the $a$-plane gives the
following transformation law for $a_D$:
$$
a_D\rightarrow a_D+a-\frac{m}{\sqrt{2}}.
$$
This transformation law can be written as the matrix acting on
the vector $(a_D,a,m)$:
$$
M=\left(
\begin{array}{ccc}
1 & 1 & -1
\\ 0 & 1 & 0
\\ 0 & 0 & 1
\end{array}
\right).
$$

What happens to the central charge after encircling singularity?
Seiberg and Witten claimed that the latter should not change under
the arisen monodromy, since it is connected to the observable mass of BPS state. The central charge invariance implies that the
row of charges must also be transformed. Apparently, it should be
rotated by the inverse monodromy matrix $M^{-1}$. Thus, we find the
charge transformation law:
\be
\label{charge transform}
(n_m,n_e,S)\rightarrow (n_m,n_e-n_m,S+n_m).
\ee

Looking at this transformation law, we notice that by multiple encircling the singularity, in the presence of a monopole, we can make the $S$-charge arbitrarily large. Let us concentrate on studying the BPS states consisted of a monopole and quarks. Quarks, being the matter fields, possess the $S$-charge. This charge simply counts the fermion number of the physical state. However, since the central charge is invariant and so does the mass or the energy of the corresponding BPS state, the number of quarks which are sitting on the same energy level, if the latter is quantized, is fixed according to the Pauli exclusive principle.\footnote{As we shall see later there is a one separate fermionic energy level in the presence of a monopole.} Thus, on the one hand the fermion number is fixed, but on the other hand it could be made arbitrarily large. This paradox may be resolved assuming that additional fermion arising after encircling the singularity leaves the quantized energy level and falls apart from the monopole.

In the limit $m\gg \Lambda$ ($g^2 \rightarrow 0$), where $\Lambda$
is the dynamically generated scale of the theory, the singularity,
corresponding to massless quark, becomes semiclassical. This means
that semiclassical approach is reliable for an investigation of
the behavior of quark-monopole bound state. Below we solve
classical equations for fermion mode in the monopole background
configuration. Studying the properties of the solution, we observe
the variation of the BPS spectrum under encircling the
singularity, which is consistent with the observations of charge
shift made above.


\section{Derivation of the fermion mode}
\subsection{Solution to Dirac equation}

Consider the $SU(2)$ theory with the action (\ref{N=1 matter}) in the semiclassical part of the moduli
 space $a\gg\Lambda$. Assume that point $a=m/\sqrt{2}$ also belong to this region. The theory contains monopoles in its spectrum. There are also solutions that correspond to matter fermions in the external monopole field. The bound state corresponds to the configuration with localized energy. Therefore the corresponding fermion mode should be normalizable. As we shall see such normalizable solutions to the classical equations exist only in certain region of the moduli space.

The fermionic components of the hypermultiplet satisfy the Dirac equation:
\be
\label{Dirac eq}
i\gm D_\mu \Psi - (\hat{m}+\sqrt{2}\hat{A})\Psi = 0,
\\
\\ \hat{m}+\sqrt{2}\hat{A}=\Re {\rm e}(m+\sqrt{2}A) +i\gamma^5\Im {\rm
m}(m+\sqrt{2}A),
\ee
where the Dirac spinor $\Psi$ is composed of two Weyl spinors as follows:
$$
\Psi=\left(
\begin{array}{c}
 \pq
\\ \bar{\psi}_{\tilde{q}}
\end{array}
\right).
$$
These two Weyl spinors belong to fundamental and antifundamental
${\cal{N}}=1$ chiral multiplets.

Working in the Gauss gauge $A_0=0$, substitute $D_0$ by the energy
eigenvalue $iE$. Now notice that the Hamiltonian commutes with the operator
$\Gamma=\gamma^0 \gamma^5$ in the case $\Im {\rm m}(m)+\sqrt{2}\Im {\rm m}(A)=0$. Thus general
solution to (\ref{Dirac eq}) could be found as the sum of $\Gamma$
eigenfunctions $\Psi=\Psi^+ +\Psi^ -$.

Substituting $\Psi^+$ instead of $\Psi$ we split the equation (\ref{Dirac eq}):
\be
\label{chiral eq}
\bigl[\sigma^k D_k -\Re {\rm e}(m+\sqrt{2}A)\bigr] \psi^+=0,
\\
\\ \bigl[ E - \Im {\rm m}(m+\sqrt{2}A)\bigr] \psi^+ =0.
\ee
Here $\psi^+$ is a chiral Weyl component of $\Psi$.

In the consideration above we treat $m$ as a real constant in contrast with
scalar field $A$ which is generally complex. However it would be more
useful for the future computations to keep $A$ real. Since there is a $U(1)_{\cal{R}}$ symmetry acting on the fields, we could choose a U(1) rotation
\begin{equation*}
\begin{array}{c}
A\rightarrow e^{2i\alpha} A,
\\
\\ m\rightarrow e^{2i\alpha} m,
\\
\\ \Psi\rightarrow e^{-i\alpha\gamma^5}\Psi,
\end{array}
\end{equation*}
under which complexity of $A$ flows into complexity of $m$. Although this
symmetry is in general anomalous, this is not the case in the semiclassical region.
Then solution to the second equation of (\ref{chiral eq}) implies that
the bound-state energy spectrum satisfies
$$E= \Im {\rm m}({\cal{M}}),$$
where ${\cal{M}}$ now defines a complex mass parameter.

In the case of real scalar field $A$ we could use the standard radially symmetric
solution for 't~Hooft-Polyakov monopole \cite{'tHooft:1974qc}
in the BPS limit \cite{BPS}\footnote{We neglect the back
reaction of matter fields to monopole fields, assuming that in semiclassical limit the
ratio of corresponding masses is small.}:
\be
\label{BPS mon}
\sqrt{2}A^i=an^i(1-F(r)),
\\
\\ \ds A^a_i=\epsilon^{aij}\frac{n^j}{r}(1-H(r)),
\ee
where $n^i$ is a unit vector, $F$ and $H$ are known functions of radius
$r$. For future convenience we chose the normalization of the scalar field $A$
in (\ref{BPS mon}) slightly different from that in \cite{SW2}. In our
conventions it has asymptotic $A\rightarrow a/\sqrt{2}$.
Furthermore, in \cite{SW2} the charges of the BPS states were renormalized in a such a way that $a\rightarrow a/2$. So our $a$ will be different from that of \cite{SW2} by the factor of $(2\sqrt{2})^{-1}$.

Substituting (\ref{BPS mon}) into (\ref{chiral eq}) we find a solution in
the form
\be
\label{solution}
(\psi^+)^{\alpha}_{~a}=\delta^{\alpha}_{~a}\chi_0\xi + n^i
(\sigma^i)^\alpha_{~\beta} \delta^{\beta}_{~a}\eta_0\zeta,
\ee
where $\alpha$ and $a$ are spinor and color indices respectively.
Functions $\chi_0$ and $\eta_0$, solving the equations at zero $m$
are defined as follows:\footnote{Here we define $\rho=\frac{r}{r_0},
\quad r_0^{-1}=a$.}
\be
\label{sol homog}
\chi_0 = \frac{1}{\sqrt{\rho {\rm sh}\rho}}{\rm th}\frac{\rho}{2},
\\
\\ \ds
\eta_0 = \frac{1}{\sqrt{\rho {\rm sh}\rho}}{\rm cth}\frac{\rho}{2}.
\ee
Functions $\xi$ and $\zeta$ satisfy the system of the first order differential
equations:
\be
\label{xieq}
\xi'=r_0m\zeta {\rm cth}^2({\rho \over 2}),
\\
\\ \ds
\zeta'=r_0m\xi {\rm th}^2({\rho \over 2}),
\ee
which implies the pair of second order equations:
\be
\label{der eq2}
\xi^{\prime\prime} + \frac{2\xi^\prime}{{\rm sh}\rho}-r_0^2m^2\xi=0,
\\
\\ \ds \zeta^{\prime\prime} - \frac{2\zeta^\prime}{{\rm sh}\rho}-r_0^2m^2\zeta=0.
\ee
Explicit solutions to this equations could be found in terms of
hypergeometric functions. We shall find normalizable solutions after an
analysis of asymptotic behavior of the general solutions to (\ref{chiral
eq}).

The solution presented above is one of positive chirality. It could be
shown as well that for negative chirality the similar solution could not be made
normalizable.
It is also natural to assume that there are no other discrete spectrum solutions to (\ref{Dirac eq}). See the discussion in \cite{HEN}.

\subsection{Analysis of the Solution}
In this subsection we investigate the asymptotic of the solution to
(\ref{Dirac eq}) and rederive the conditions under which this solution exists
\cite{HEN}. Using the asymptotic we also derive the explicit solution in
terms of hypergeometric functions.

Introduce a pair of functions $\chi,\eta$ that will characterize the
asymptotic behavior of solution (\ref{solution}):
$$
\chi=\chi_0\xi, \qquad \eta=\eta_0 \zeta.
$$
Since we are interested in soliton-fermion bound states,
the fermion mode we are looking for should be a normalizable one, i.e.\ it
should decrease fast enough at infinity and be regular at the origin,
namely
$$\chi(0)={\rm const},\qquad \eta(0)=0.$$
In the $\rho \rightarrow 0$ limit the fermion mode behaves as the solution
to the equations
\be
\label{zero asympt1}
\xi^{\prime\prime} + \frac{2\xi^\prime}{\rho}-r_0^2m^2\xi=0,
\\
\\ \ds \zeta^{\prime\prime} - \frac{2\zeta^\prime}{\rho}-r_0^2m^2\zeta=0.
\ee
The regular solutions to (\ref{zero asympt1}) are as follows:
\be
\label{zero asympt2}
\xi=C\frac{{\rm sh}(r_0m\rho)}{\rho},
\\
\\ \zeta = C(r_0m\rho-1)e^{r_0m\rho}+ C(r_0m\rho+1)e^{-r_0m\rho},
\ee
with the coefficients fixed by asymptotic behavior of (\ref{xieq}).

Assume that
we have a solution that is regular at the origin. From the asymptotic
(\ref{zero asympt2}) it follows that the first derivative of this solution
is zero and the second derivative is positive at this point. Furthermore,
the second derivative in exact equations (\ref{der eq2}) is positive wherever
first derivative vanishes. Therefore such solutions could have only minima
and cannot be normalizable.
This means that
solution to (\ref{der eq2}) is either regular at the origin and divergent at
infinity or regular at infinity and divergent at the origin. However, if
we look at (\ref{sol homog}), we shall see that under certain conditions
we could construct a normalizable zero-mode. The regular at the origin and
increasing at infinity solution to (\ref{der eq2}) has the asymptotic
$$
\xi,\zeta \sim e^{r_0m\rho}.
$$
It follows from (\ref{sol homog}) that the functions
$\chi_0$,$\eta_0$ have the asymptotic
$$
\chi_0, \eta_0 \sim \frac{e^{-\rho}}{\sqrt{\rho }},
$$
and that for the functions $\chi$,$\eta$:
$$
\ds \chi,\eta \sim \sqrt{\frac{1}{\rho }}\exp[{(r_0m-\frac{1}{2})\rho}].
$$
We see that for the existence of normalizable zero mode the following conditions
should be satisfied:
$$
2r_0 m < 1, \quad {\rm or} \quad \frac{2m}{a}< 1,
$$
or $a>m/\sqrt{2}$ in the standard normalization of \cite{SW2}. This result was
originally derived in \cite{HEN}, and then was confirmed by CMS considerations
in \cite{VainsteinR} in the weak coupling limit.

Now turn to the equations (\ref{der eq2}). Changing variables from
$\rho$ to $x=({\rm ch}\rho+1)/2$ we obtain a pair of hypergeometric
equations:
\be
\label{hypgeom2}
x(x-1)y^{\prime\prime}_1+(x+\frac{1}{2})y_1^\prime-r_0^2m^2y_1=0,
\\
\\ \ds x(x-1)y^{\prime\prime}_2+(x-\frac{3}{2})y_2^\prime-r_0^2m^2y_2=0.
\ee
Solving (\ref{hypgeom2}) and expressing the solution in terms of original
variable $\rho$ we find the unknown functions $\xi$ and $\zeta$:
\begin{equation*}
\begin{array}{c}
\ds \xi(\rho)=CF(\frac{m}{a},-\frac{m}{a},-\frac{1}{2},{\rm ch}^2\frac{\rho}{2}),
\\
\\ \ds \zeta(\rho)= \tilde{C}
F(\frac{m}{a},-\frac{m}{a},\frac{3}{2},{\rm ch}^2\frac{\rho}{2}).
\end{array}
\end{equation*}
Here we took into account the asymptotic of solution determined above.
The asymptotic of (\ref{xieq}) also fixes the relation $C=\tilde{C}$.

\subsection{BPS State Decay}
>From the considerations above we found that normalizable fermion mode with
the energy defined by complex mass parameter $E=\Im {\rm m}{\cal{M}}$ exists only
in the region of moduli space defined by\footnote{In the fermion equations $a$
was treated as $vev$ of real field $A$. To define this condition for all complex
moduli plane we substitute the latter by absolute value of complex coordinate
$|a|$.}
\be
\label{norm cond2}
\Re {\rm e}{\cal{M}}=m<\frac{|a|}{2}.
\ee
Now for complete understanding of the process we need one more thing, namely
to find what region of the moduli space corresponds to the solutions to Dirac
equation (\ref{Dirac eq}) from continuous spectrum, i.e.\ non-bound BPS
states. Considering asymptotic of (\ref{Dirac eq}) at infinity we obtain the
system of linear differential equations, neglecting exponentially decaying
non-constant coefficients. The corresponding characteristic equation gives
for fixed energy $E$
$$
E=\pm\left|m-\frac{a}{2}\right|.
$$

This condition defines a cone in the $E-a$, or equivalently in the $E-{\cal{M}}$
parameter space, with the singularity at $a=2m$ (Fig.1). Further we substitute the
investigation on the complex $a$ plane by that on the $\cal{M}$ plane, which
is equivalent due to $U(1)_{\cal{R}}$ symmetry. Continuous spectrum
belongs to the interior of this cone. The region of discrete spectrum is
an inclined plane $E=\Im {\rm m}{\cal{M}}$, which is bounded due to the condition (\ref{norm
cond2}).

\begin{figure}[]
\epsffile{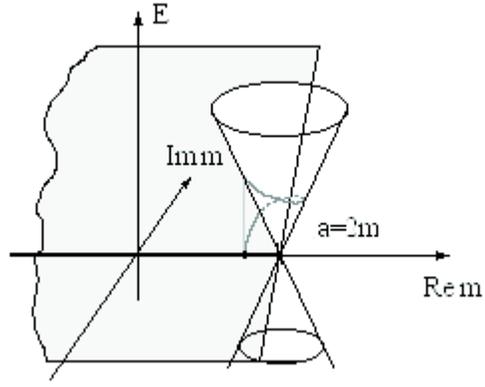}
\caption{The region of discrete spectrum (inclined semiplane) and the region of continuous spectrum (interior of the cone) intersect via the line $E=\Im {\rm m}
{\cal{M}}$
which goes through the point $a=2m$.}
\end{figure}

Now a closed contour on the complex $\cal{M}$ plane corresponds to some
3-dimensional curve, which is restricted to belong either to inclined plane of
discrete spectrum or to the interior of the continuous spectrum cone, where,
we suppose, are the only possible loci for consistent physical state. Since
the contour must be closed and smooth, this curve must have a spiral form.
This means that encircling the singular point we cannot return to the
initial value of energy. Furthermore, the final point of a spiral curve
belongs to the state from continuous spectrum, which means a decay of the
initial bound state.

To clarify what happens when we encircle the singularity, consider the
dependence of energy $E$ on the angle of rotation $\varphi$ in the
$\cal{M}$ plane (Fig.2). We start at some real mass value $m_0$ and hence at zero
energy.
At $\varphi=\pi/2$ we reach the continuous spectrum. For $\varphi$
from $\pi/2$ to $3\pi/2$ there is no any bound state, i.e. the
initial bound state has decayed. After a full rotation $\varphi=2\pi$ the
mode from continuous spectrum cannot descend to discrete spectrum and the only
allowed values of energy will be $E\geq |m_0-a/2|$. However, if we
move in the opposite direction, starting from $\varphi=2\pi$, we shall also come
to continuous spectrum at $\varphi=3\pi/2$, but in the part of it which
is below zero level. Since we move adiabatically, and all the states for fermion below
zero are occupied, we conclude that after encircling the singularity one fermion
mode runs to continuous spectrum, and another one comes from the Dirac sea below.

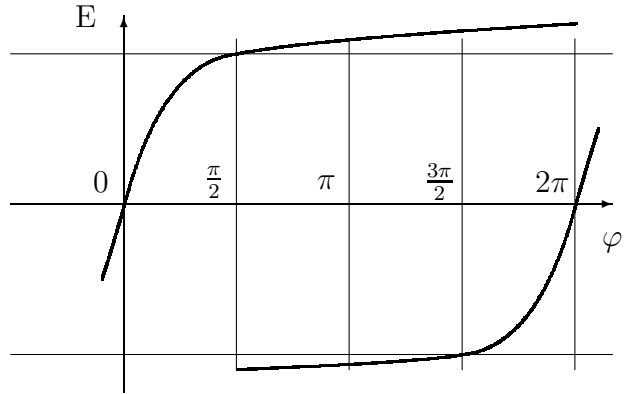
\begin{figure}[h]
\label{Fig2}

\special{em:linewidth 0.4pt}
\unitlength 1mm
\linethickness{0.4pt}
\begin{picture}(150,55.00)(8,0)
\put(120,20){\makebox(0,0)[cc]{$\varphi$}}
\put(50,50){\makebox(0,0)[cc]{E}}

\emline{40}{45.00}{1}{120}{45}{2}
\emline{40}{5.00}{3}{120}{5}{4}
\emline{70}{3.00}{5}{70}{47}{6}
\emline{85}{3.00}{7}{85}{47}{8}
\emline{100}{3.00}{9}{100}{47}{10}
\emline{115}{3.00}{11}{115}{47}{12}

\put(52,28){\makebox(0,0)[cc]{0}}
\put(67,28){\makebox(0,0)[cc]{$\frac{\pi}{2}$}}
\put(82,28){\makebox(0,0)[cc]{$\pi$}}
\put(97,28){\makebox(0,0)[cc]{$\frac{3\pi}{2}$}}
\put(112,28){\makebox(0,0)[cc]{$2\pi$}}

\put(40,25){\vector(1,0){80}}
\put(55,0){\vector(0,1){50}}

\linethickness{0.8pt}
\bezier{150}(55,25)(60,44)(70,45)
\bezier{150}(100,5)(110,6)(115,25)
\bezier{300}(70,45)(80,47)(115,49)
\bezier{200}(100,5)(95,4)(70,3)
\bezier{150}(55,25)(53,18)(52,15)
\bezier{150}(118,35)(117,32)(115,25)

\end{picture}
\caption{The initial bound state decays after crossing the point of
marginal stability
at $\varphi = \pi/2$. Another fermion mode replaces it rising from the
continuous spectrum below.
}
\end{figure}

\section{Conclusions}

Above we have found and investigated solutions to the classical equations of
motion (\ref{Dirac eq}). We were searching for the solutions with localized
energy or, in the other words, normalizable solutions. We have established
that this kind of field configuration exists only in the definite region of
the moduli space. Our interest was in process of encircling a singular
point. We have found that starting from the bound state and closing a loop
around the singularity, one comes to the multiparticle state, consisted of
the bound state of fermion and monopole, and the distant delocalized fermion
mode in the continuous part of the energy spectrum.

This result was expected from Seiberg and Witten considerations of moduli
space of the ${\cal{N}}=2$ theory. Considering just the BPS mass formula
(\ref{Matter CC}) and claiming the invariance of the central charge under
monodromy it was stated that the charges should transform as follows:
$$(n_m,n_e,S)\rightarrow (n_m,n_e-n_m,S+n_m).$$
This is indeed the case in semiclassical analysis with charges (1,-1,1).
Namely the magnetic charge is conserved, the electric charge is shifted
by one, due to the fermion mode in continuous spectrum. The $S$-charge also accounts this additional fermion mode.

What is the physical meaning of this result? Strictly speaking we do not discuss the process of some observable decay of a physical state. The original word {\it{decay}} may sound slightly misleading, because the matter concerns rather {\it{delocalization}} than decay. The process of encircling singularity does not seem to be very physical, since the motion around singular point takes place in the abstract moduli space. However, there is a strong evidence that the moduli spaces of various supersymmetric theories are related to the phase spaces of the certain dynamical systems. (For discussion and references see \cite{Morozov}.) Not taking into account this relation, the motion around the singularity actually corresponds to the duality transformation, or transition from one theory description to another. We are familiar to the special cases of such transition.

Consider the electric-magnetic duality. According to the Dirac quantization
rule, the large value of electric charge $e$ corresponds to the small value
of magnetic charge $1/e$.  When electric charge becomes large, it is
nonsensical to use the perturbative expansion with electric charge as the
parameter. The dual theory with the small magnetic charge may be used instead. In
fact we are still working with the same theory, but use another description
for it. According to this description, the fields which were electrically
charged, look magnetically charged after the transition.

In the above example we consider the simple duality transformation. From the
point of view of Seiberg-Witten theory the latter case corresponds to the
real scale (one-dimensional moduli space) and only a subgroup of the duality
group. In this paper we worked with the complex moduli space, and therefore, the complex
scale. Due to the complexity the multivaluedness appeared. This
multivaluedness has naturally described the existing variety of descriptions of the theory on a complex moduli space. We have been convinced that for the case of ${\cal{N}}=2$ SYM theory with matter, the whole set of charges $(n_m,n_e,S)$ is necessary to take into account, for correct application of duality transformations.

\vspace{0.5cm}
We are grateful to A.Gorsky and especially to Kostya Selivanov for initiating
this work, for various support and useful discussions at the different stages.
We also acknowledge to E.~Ahmedov, S.~Dubovsky, V.~Pestun, A.~Solovyov, A.~Vainshtein and K.~Zarembo for clarifying several subtle questions.
A.D. would like to thank the Carg\`ese 2002 ASI, where the part of work
was done. Analogously D.M. would like to thank the Department of
Theoretical Physics at Uppsala University for hospitality.

The work was supported in part by grants  RFBR 01-01-00549, INTAS
00-334 (A.D.) and  INTAS 00-561 (D.M.).

\end{document}